\documentclass[12pt,letterpaper,twocolumn,british,sort,prl,reprint]{revtex4-1}

\usepackage[T1]{fontenc}
\usepackage[utf8]{inputenc}
\usepackage{color}
\usepackage{textcomp}
\usepackage{graphicx}

\makeatletter

 \@ifundefined{textcolor}{}
 {%
   \definecolor{BLACK}{gray}{0}
   \definecolor{WHITE}{gray}{1}
   \definecolor{RED}{rgb}{1,0,0}
   \definecolor{GREEN}{rgb}{0,1,0}
   \definecolor{BLUE}{rgb}{0,0,1}
   \definecolor{CYAN}{cmyk}{1,0,0,0}
   \definecolor{MAGENTA}{cmyk}{0,1,0,0}
   \definecolor{YELLOW}{cmyk}{0,0,1,0}
 }

\usepackage[cbgreek]{textgreek}
\usepackage{hyperref}

\makeatother

\usepackage{babel}
\begin{document}

\title{The mechanism of caesium intercalation of graphene}

\author{{\small \vphantom{}M. Petrović$^{1}$, I. Šrut Rakić$^{1}$, S.
Runte$^{2}$, C. Busse$^{2}$, J. T. Sadowski$^{3}$, P. Lazić$^{4}$,
I. Pletikosić$^{1}$, Z.-H. Pan$^{5}$, M. Milun$^{1}$, P. Pervan$^{1}$,
N. Atodiresei$^{6}$, R. Brako$^{4}$, D. Šokčević$^{4}$, T. Valla$^{5}$,
T. Michely$^{2}$ and M. Kralj$^{1,}$}}

\email{mkralj@ifs.hr}

\affiliation{{\footnotesize \vphantom{}$^{1}$ Institut za fiziku, Bijenička
46, 10000 Zagreb, Croatia}}

\affiliation{{\footnotesize $^{2}$ II. Physikalisches Institut, Universität zu
Köln, Zülpicher Str. 77, 50937 Köln, Germany}}

\affiliation{{\footnotesize $^{3}$ Center for Functional Nanomaterials, Brookhaven
National Lab, Upton, New York 11973, USA}}

\affiliation{{\footnotesize $^{4}$ Institut Ruđer Bošković, Bijenička 54, 10000
Zagreb, Croatia}}

\affiliation{{\footnotesize $^{5}$ Condensed Matter Physics \& Materials Science
Department, Brookhaven National Lab, Upton, New York 11973, USA}}

\affiliation{{\footnotesize $^{6}$ Peter Grünberg Institut \& Institute for Advanced
Simulation, Forschungszentrum Jülich and JARA, 52425 Jülich, Germany}}
\begin{abstract}
Properties of many layered materials, including copper- and iron-based
superconductors, topological insulators, graphite and epitaxial graphene
can be manipulated by inclusion of different atomic and molecular
species between the layers via a process known as intercalation. For
example, intercalation in graphite can lead to superconductivity and
is crucial in the working cycle of modern batteries and supercapacitors.
Intercalation involves complex diffusion processes along and across
the layers, but the microscopic mechanisms and dynamics of these processes
are not well understood. Here we report on a novel mechanism for intercalation
and entrapment of alkali-atoms under epitaxial graphene. We find that
the intercalation is adjusted by the van der Waals interaction, with
the dynamics governed by defects anchored to graphene wrinkles. Our
findings are relevant for the future design and application of graphene-based
nano-structures. Similar mechanisms can also play a role for intercalation
of layered materials.
\end{abstract}
\maketitle
Rapid advance in epitaxial growth has resulted in uniform large-area
high-quality epitaxial graphene (e-Gr) \cite{Emtsev2009-1,Wintterlin2009-1,Li2009-1},
a prerequisite for future applications of graphene \cite{Bae2010-1,Lin2011-1}.
In search for novel properties, e-Gr systems are often modified with
different atoms or molecules intercalated between graphene and its
support—for example, to decouple graphene from its substrate by reducing
the bonding interaction \cite{Riedl2009-1,Varykhalov2008-1}. One
important aspect of intercalation is chemical doping of graphene sheets
\cite{Liu2011}, an effect similar to electric field doping in field-effect
graphene transistors \cite{Novoselov2004-1}. Compared to field induced
doping, chemical doping has the additional options to open a band
gap at the Dirac point \cite{Jung2009-1} or even shift the the Fermi
level up to the extended \textpi {*} band van Hove singularity of
graphene \cite{McChesney2010-1}. 

The diversity and sensitivity of the physical and chemical properties
of e-Gr \cite{Liu2011}, graphite \cite{Dresselhaus1981,Chung2002}
and other layered systems \cite{Valla2012} with respect to intercalation
of charge-donating species is directly related to prominent effects,
such as superconductivity in graphite \cite{Weller2005-1,Emery2005-1}
and the uncharted one in graphene \cite{Profeta2012-1,Uchoa2007-1}.
From the view point of chemical kinetics, understanding the penetration
and diffusion of ions under graphene sheets in atomistic detail is
of fundamental importance for the design of novel batteries and supercapacitors
\cite{Yoo2011-1}. There are other potentially useful properties associated
with intercalation of e-Gr systems. For instance, it was demonstrated
that by precise control of the intercalation interface, laterally
well-defi{}ned mesoscopic regions of \textit{n}- and \textit{p}-doped
graphene—that is, \textit{p-n} graphene junctions, can be formed \cite{Emtsev2011}.
It was also demonstrated that it is possible to form well-defined
ferromagnetic nano-islands under e-Gr \cite{Sicot2012-1}. Therefore,
it is important to understand in detail how the properties of chemically
modified graphene depend on its chemical environment. For example,
for spintronic applications, on the one hand intercalation of ferromagnetic
3d-metals such as Co \cite{Decker2013} or Ni \cite{Pacile2013,Sicot2012-1}
is explored but it results in a strong chemical interaction between
the metal d-orbitals and \textpi -bands of graphene. On the other
hand, a more promising approach to design magnetic properties of graphene
by using its almost unperturbed \textpi -bands is the intercalation
o\textcolor{black}{f a wider class of systems of the same nature—such
as, alkali metals, alkaline earth metals o}r lanthanides (for example,
Nd or Eu that contain 4f- and 6s-electrons) \cite{Schumacher2013-1}.

Graphene on Ir(111), e-Gr/Ir, is recognized for its unique structural
quality \cite{vanGastel2009-1} and weak bonding to the substrate
\cite{Busse2011-1} that results in an electronic structure resembling
that of freestanding graphene \cite{Pletikosic2009-1,Kralj2011-1}.
Despite macroscopic uniformity, it was shown that different metals
and molecules can be successfully intercalated between graphene and
iridium \cite{Bianchi2010-1,Pletikosic2012-1,Tontegode1991-1,Granas2012-1,Larciprete2012-1,Rutkov1995-1,Vinogradov2013-1,Schumacher2013-1,Decker2013,Pacile2013}.
These findings not only call for an explanation on how the intercalation
can occur through an apparently perfect graphene membrane but also
for the mechanism driving the intercalation.

Here we characterize the process of adsorption and intercalation of
Cs in microscopic detail by utilizing \textit{in situ} low energy
electron microscopy (LEEM), scanning tunneling microscopy (STM), low
energy electron diffraction (LEED), angle-resolved photoemission spectroscopy
(ARPES), and density functional theory (DFT) calculations. For comparison
we characterize also the intercalation of Li, which has a much smaller
ionic radius than Cs. We describe the energetics and the novel mechanisms
which explain the kinetics of intercalation and entrapment of alkali
atoms under epitaxial graphene. We find that these processes are regulated
by the van der Waals interaction, with defects at graphene wrinkles
acting as penetration site\textcolor{black}{s for alkali atoms. }The
reported mechanism may be specific to the particular system studied.
Nevertheless, we believe that our study represents a very important
and necessary step in gaining a fundamental knowledge on the s-\textpi{}
interactions and the intercalation process itself. This information
can be further used to unveil the properties of more complicated intercalated
systems—for example, where additional complex spin-dependent interactions
take place.

\section{Results}

\subsection{Characterization of the adatom and intercalated phases}

The experiments presented below indicate the formation of two phases
upon Cs deposition on e-Gr/Ir. After the initial dilute-phase on graphene,
the \textalpha -phase, reaches its maximum density at about 0.06
monolayer (ML), additionally deposited Cs is intercalated and forms
the \textgamma -phase as a ($2$\texttimes{}$2$) structure. Upon
further deposition the area fraction of the \textgamma -phase increases
at the expense of the area fraction of the \textalpha -phase. Once
the entire sample is transformed to the \textgamma -phase, it further
becomes denser through the applied Cs pressure until the saturation
density is eventually reached, when the \textgamma -phase has the
($\sqrt{3}$\texttimes{}$\sqrt{3}$)R30\textdegree{} structure relative
to Ir(111) at a coverage taken as a definition of 1 ML of Cs.

\begin{figure}
\begin{centering}
\includegraphics{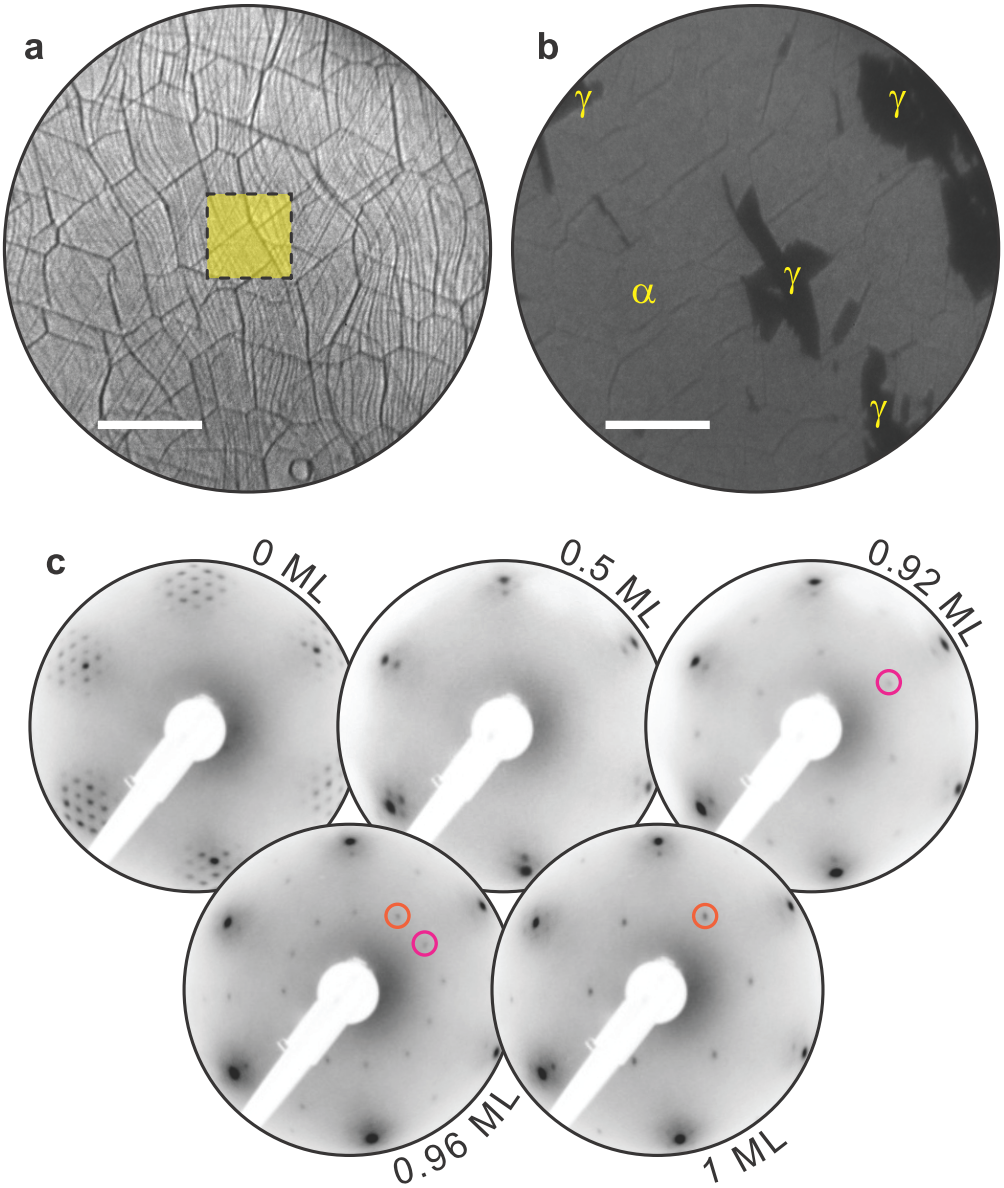}
\par\end{centering}

\caption{\textbf{LEEM and LEED characterization.} (\textbf{a}) LEEM topograph
of a uniform e-Gr monolayer on Ir(111) showing characteristic surface
features: graphene wrinkles and substrate steps. The yellow square
indicates the area analyzed for the dynamics of intercalation (cf.
Fig. 5). (\textbf{b}) Same area as in \textbf{a} but after deposition
of 0.2 monolayer (ML) of Cs. A moderate decrease (\textalpha -phase,
marked as \textalpha ) and a strong decrease (\textgamma -phase,
marked as \textgamma ) in reflectivity are visible. Scale bars in
\textbf{a} and \textbf{b} correspond to 3 \textmu m. (\textbf{c})
LEED patterns of structures obtained for various amounts of deposited
Cs (indicated for each pattern). The 0 ML pattern was recorded at
117 eV and all remaining patterns at 130.5 eV electron energy. Magenta
circles indicate a ($2$\texttimes{}$2$) structure relative to e-Gr
while orange circles indicate a ($\sqrt{3}$\texttimes{}$\sqrt{3}$)R30\textdegree{}
structure relative to Ir.}

\end{figure}

Fig. 1 shows the fingerprints of structures being formed in the process
of intercalation. The corresponding LEEM image of the initial e-Gr/Ir
surface (Fig. 1a) reveals several structural features. Thick dark
lines are wrinkles—graphene areas delaminated from the substrate—a
signature of strain release in graphene \cite{Hattab2012-1,NDiaye2009-1}.
In addition, steps of the Ir(111) substrate are resolved which wiggle
mostly vertically as well as diagonally (several straight glide steps)
through the image. The LEEM images were recorded for various amounts
of deposited Cs, up to 1 ML. Fig. 1b shows a LEEM micrograph of the
surface after deposition of 0.2 ML of Cs. Dark regions with sharp
boundaries are visible and attributed to the \textgamma -phase. The
remaining homogeneous surface contrast is attributed to the \textalpha
-phase. For more details on LEEM characterization see Supplementary
Note 1 and Supplementary Movie 1. In addition to LEEM imaging at fixed
electron energy, we also performed the \textit{IV}-LEEM measurements
before and after deposition of 0.2 ML of Cs. The characteristic reflectivity
curves from pristine e-Gr/Ir, the \textalpha -phase and the \textgamma
-phase areas are shown in Supplementary Fig. S1. The origin of apparent
contrast differences in Fig. 1a and 1b\textcolor{black}{{} is clearly
visible }in the reflectivity curves at the imaging energy of 1.5 eV.
The \textgamma -phase reflectivity shows a dip with a minimum close
to that energy which is a clear indication of the formation of a new
dense layer \cite{Chung2003}.

\begin{figure}
\begin{centering}
\includegraphics{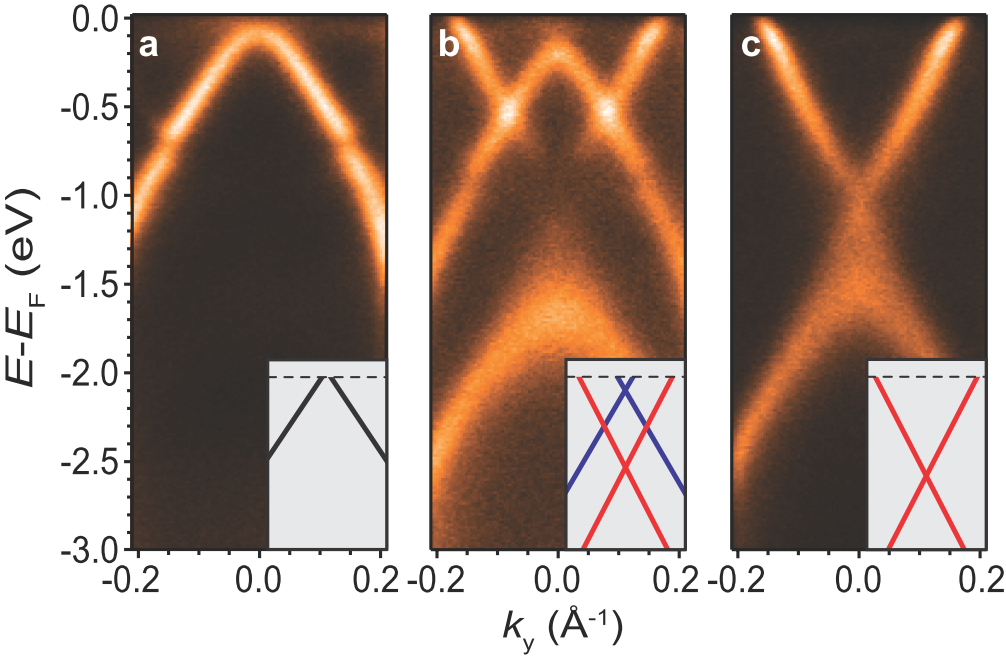}
\par\end{centering}

\caption{\textbf{ARPES characterization.} ARPES maps (vertical axis: binding
energy $E-E_{\text{F}}$ referenced to the Fermi level $E_{\text{F}}$,
horizontal axis: wave vector $k_{\text{y}}$) taken across the K point
of graphene for different amounts of deposited Cs: (\textbf{a}) pristine
graphene, (\textbf{b}) 0.5 ML and (\textbf{c}) 1 ML of Cs. Insets
in \textbf{a}-\textbf{c} schematically indicate Dirac cones visible
in ARPES maps, corresponding to pristine (black), light (blue) and
heavily (red) electron doped Dirac cones.}
\end{figure}

Fig. 1c shows the emerging structures observed by LEED, recorded at
room temperature. The diffraction spots of e-Gr/Ir before deposition
reveal the characteristic moiré pattern of uniformly aligned graphene.
For coverages below 0.9 ML, no clear Cs superstructure is observed
in LEED (cf. 0.5 ML). Apparently, at room temperature intercalation
islands of finite size do not exhibit a stable crystal structure.
This was further corroborated by a micro-LEED imaging in the LEEM
setup (not shown). Only at high coverages of $\sim$0.9 ML and above,
or at low temperatures, well defined superstructures of intercalated
Cs are observed. In the range from 0.9 to 1 ML of Cs we find in LEED:
a ($2$\texttimes{}$2$) structure relative to e-Gr (for 0.92 ML),
a mixture of a ($2$\texttimes{}$2$) relative to e-Gr and a ($\sqrt{3}$\texttimes{}$\sqrt{3}$)R30\textdegree{}
structure relative to Ir (for 0.96 ML) and finally for 1 ML a ($\sqrt{3}$\texttimes{}$\sqrt{3}$)R30\textdegree{}
pattern relative to Ir. In addition, the relevant superstructures
are presented in Supplementary Table S1.

The electronic structure of e-Gr/Ir in Fig. 2a is characterized by
a sharp Dirac cone, exhibiting slight \textit{p}-doping of graphene,
consistent with previous experiments \cite{Pletikosic2009-1,Kralj2011-1}.
The deposition of Cs leads to the following modifications of the ARPES
spectra; for low Cs coverage, of up to $\sim$0.06 ML, the Dirac cone
of slightly \textit{p}-doped e-Gr/Ir shifts progressively to higher
binding energies as the Cs coverage is increased. The Dirac point
shifts down to 0.2 eV below the Fermi level, making graphene slightly
\textit{n}-doped. From the ARPES spectra we obtain that the maximum
electron doping in the \textalpha -phase is \emph{n} $\approx$ 4$\times$10$^{12}$
cm$^{-2}$ corresponding to a charge transfer of 0.11 electrons from
each Cs atom in the \textalpha -phase to graphene (see Supplementary
Fig. S2). For more than $\sim$0.06 ML, an additional Dirac cone appears
at higher binding energy due to phase separation; Fig. 2b corresponds
to 0.5 ML of Cs and it shows two Dirac cones, light and heavily electron
doped, corresponding to the the \textalpha -phase and the \textgamma
-phase, respectively. The \textalpha -phase cone decreases in intensity
and completely disappears above $\sim$0.9 ML of Cs and only the \textgamma
-phase cone remains visible. Once the entire sample is covered by
the \textgamma -phase, additional deposition from 0.9 ML to 1 ML
(Fig. 2c) increases its density, consistent with the evolution of
the LEED patterns and the additional shift of the Dirac point (see
Supplementary Fig. S3 and Supplementary Note 2). The maximum doping
of \emph{n} $\approx$ 1$\times$10$^{14}$ cm$^{-2}$ implies a charge
transfer of 0.20 electrons per Cs atom to graphene.

\begin{figure}
\begin{centering}
\includegraphics{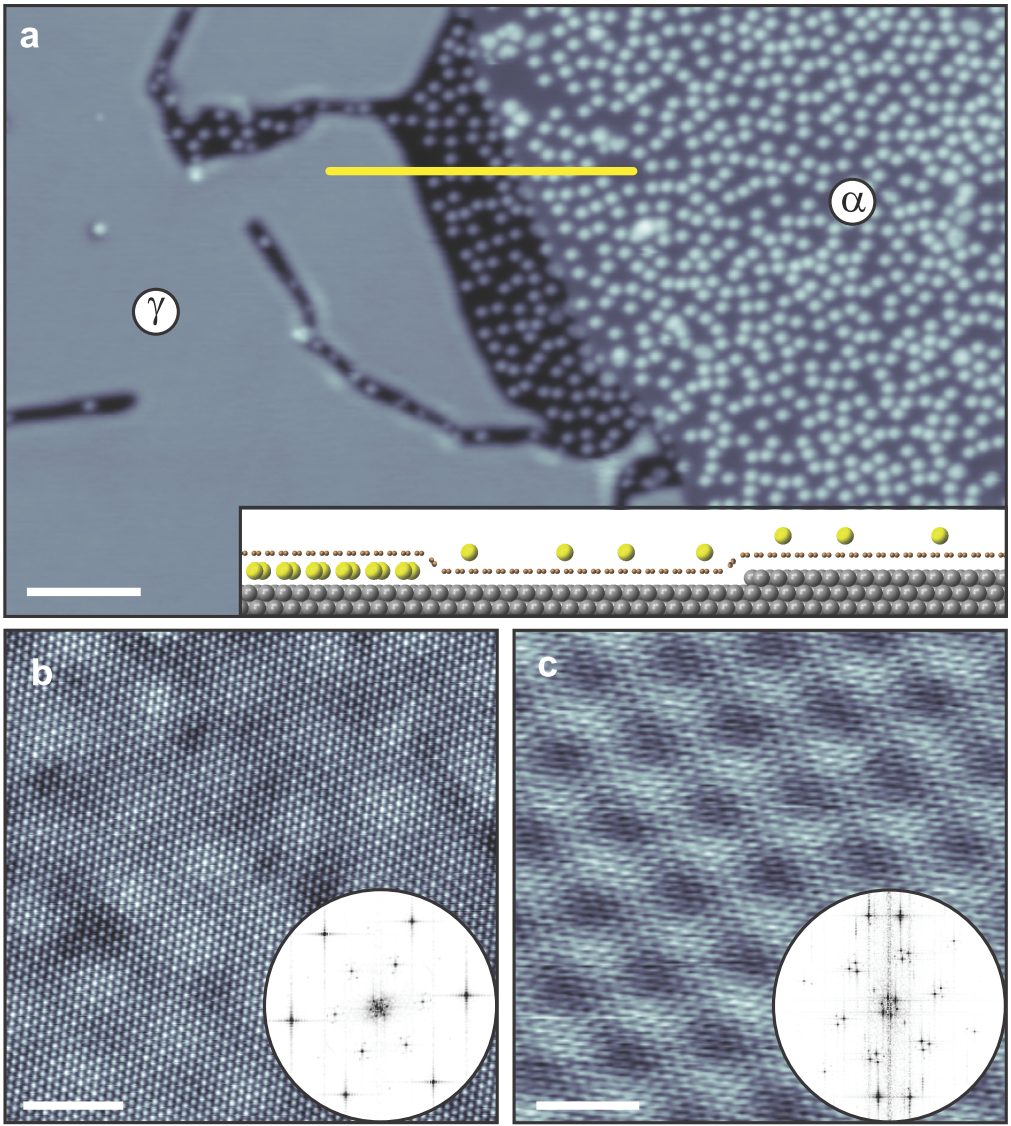}
\par\end{centering}

\caption{\textbf{STM characterization.} (\textbf{a}) An STM topograph showing
two characteristic areas found after deposition of 0.5 ML of Cs: the
adatom \textalpha -phase (marked as \textalpha ) and the intercalated
\textgamma -phase (marked as \textgamma ). The inset shows a schematic
illustration of the topographic profi{}le across different regions
of the STM image indicated by a yellow horizontal line. White scale
bar corresponds to 10 nm. (\textbf{b},\textbf{c}) High-resolution
STM topographs of two different structures revealed in the \textgamma
-phase areas: ($2$\texttimes{}$2$) relative to graphene in \textbf{b}
and ($\sqrt{3}$\texttimes{}$\sqrt{3}$)R30\textdegree{} relative
to Ir in \textbf{c}. The circular insets in \textbf{b} and \textbf{c}
show the corresponding Fourier transforms, the scale bars correspond
to 3 nm. All images were recorded at 6 K.}

\end{figure}

\begin{figure*}
\begin{centering}
\includegraphics[width=18cm]{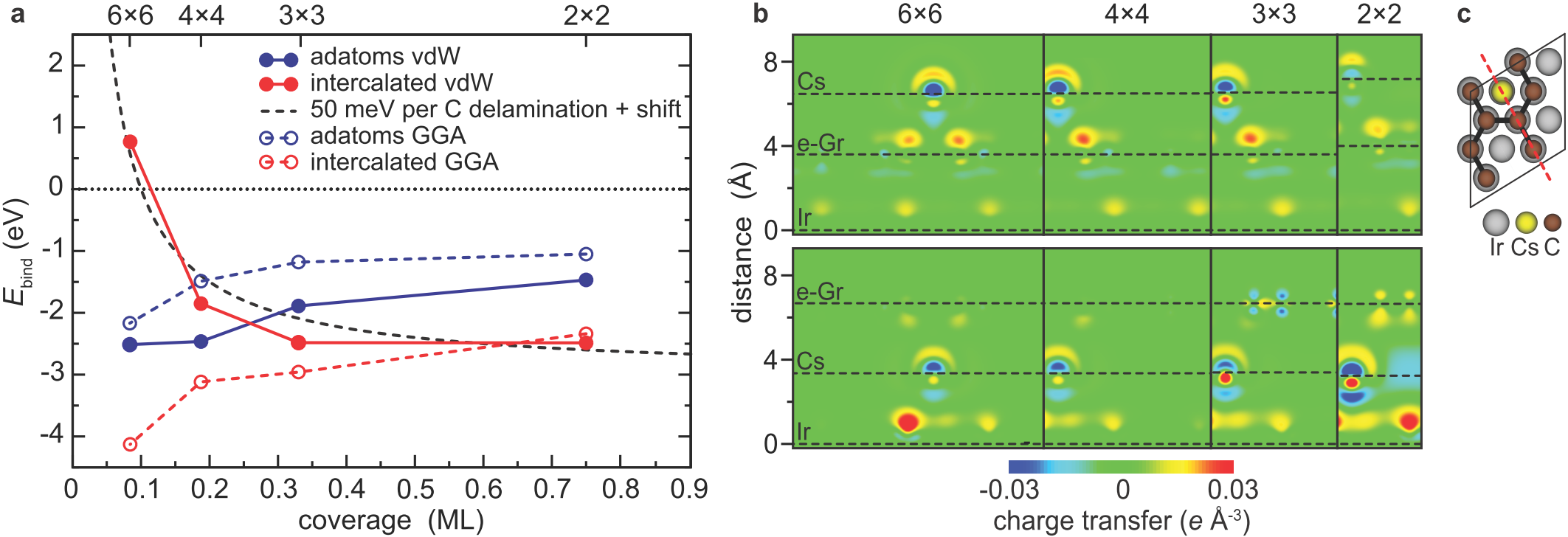}
\par\end{centering}

\caption{\textbf{Binding energies and charge transfer calculations.} (\textbf{a})
Binding energies ($E_{\text{bind}}$) as a function of Cs coverage
in monolayers (ML) for the intercalated and the adatom phase. The
plotted values are given per unit cell used in the calculations and
referenced to the corresponding pristine e-Gr/Ir system plus a free
Cs-atom. The open symbols and dashed lines show the results obtained
using the generalized gradient approximation (GGA) functional while
the full symbols and full lines are obtained with inclusion of the
non-local van der Waals (vdW) correlation. The dashed black line is
a guide to the eye simulating a 50 meV delamination energy per C atom
plus a constant offset. (\textbf{b}) Charge transfer (colour-coded
in units of electrons per Å$^{3}$) in the adatom phase (upper row)
and in the intercalated phase (lower row). The corresponding Cs concentrations
are marked on top of the panel. Black dashed lines indicate the positions
of Ir, e-Gr and Cs layers. (\textbf{c}) A model indicating the position
of a geometric plane (dashed red) across which charge transfer in
\textbf{b} is plotted.}
\end{figure*}

Additional insight comes from STM. After deposition of 0.5 ML Cs at
room temperature and cooling the sample to 6 K prior to STM measurements
to enable stable imaging, two characteristic areas can be distinguished
in the topographs. As shown in Fig. 3a, we identify the area with
disordered bright protrusions corresponding to Cs adatoms as the \textalpha
-phase. The concentration of adatoms varies slightly over the sample
surface with the maximum corresponding to hexagonally arranged atoms
at a nearest neighbor distance of (1.86 $\pm$ 0.01) nm—that is, 0.063
ML. The flat area of Fig. 3a is apparently 0.4 nm higher than its
surrounding, indicative for the presence of an additional dense layer,
which we attribute to the \textgamma -phase. Finally, note straight
and thin channels in the intercalation mesa, which we relate to the
inhomogeneous binding of graphene to the substrate marking areas with
comparatively high delamination energy \cite{Schumacher2013-1}. A
more detailed STM characterization of the \textgamma -phase is possible
through high-resolution STM imaging. The STM topographs in Fig. 3b
and 3c show atomic resolution obtained while scanning over two different
intercalated areas. In both images one can resolve the graphene unit
cell as well as a weak moiré superstructure. This is also visible
in the Fourier transforms of the STM images shown in Fig. 3b and 3c
insets. A closer inspection of the images and their Fourier transforms
enables an identification of the two different underlying Cs superstructures:
the ($2$\texttimes{}$2$) structure relative to graphene in Fig.
3b and the ($\sqrt{3}$\texttimes{}$\sqrt{3}$)R30\textdegree{} structure
relative to iridium in Fig. 3c. The STM findings are thus fully in
agreement with LEED characterization.

\subsection{DFT calculations}

The main result of our DFT calculations is shown in Fig. 4a, where
the binding energies for different Cs coverages corresponding to the
adatom and intercalated phases are plotted. Calculations with the
van der Waals (vdW) interaction included show a clear crossover between
the adatom and intercalated phases for dilute and dense concentrations,
respectively. For considered Cs concentrations, the charge transfer
in the adatom and the intercalated phase is visualized in Fig. 4b.\textcolor{black}{{}
Looking at the charge transfer alone,} the main difference between
the intercalated and the adatom structure comes from the much larger
amplitudes of charge rearrangement in the intercalated system. There,
the Cs atom causes a large redistribution of charge on both, the Ir
surface and the graphene (lower row of Fig. 4b), mimicking the electrostatic
glue for floating \textit{n}-doped graphene. In contrast, in the adatom
case most of the charge rearrangement takes place on the graphene
sheet close to Cs (upper row of Fig. 4b). In order to determine the
importance of the vdW interaction, in Fig. 4a we also present the
results obtained using a semi-local functional which does not contain
the vdW correlation effects. This functional fails to correctly describe
the experimental observations (see below and also Supplementary Fig.
S4 and S5 and Supplementary Note 3).

\subsection{Intercalation and de-intercalation dynamics}

In addition to the phases described above, LEEM sequences taken during
Cs deposition and desorption provide information on the kinetics of
these processes (see Supplementary Note 1 and 4 and also Supplementary
Movie 1 and 2). The analysis of such sequences shows that the growth
of the \textgamma -phase is discontinuous, composed of sudden and
rapid advancements followed by extended periods of stagnation. This
is visualized in Fig. 5 where snapshots of the same sample area at
the indicated incremental times are displayed. An area of pristine
e-Gr/Ir is shown in Fig. 5a. In Fig. 5b the same area is shown but
with the network of wrinkles (full lines) and steps (dotted lines).
They divide the surface into an assembly of small tetragonal micro-tiles.
We find that these features are essential for the evolution of the
intercalated phase. In particular, the \textgamma -phase always forms
next to wrinkles, typically at crossing points of wrinkles. Virtually
at any time, all boundaries of such intercalated islands are identified
as wrinkles or surface steps (Fig. 5c-5i). Hence their expansion can
be described as tiling. For the characteristic surface area studied
in our work, the density of wrinkle- and step-line substructures differed
almost by an order of magnitude, 1 \textmu m$^{-1}$ for wrinkles
and 7 \textmu m$^{-1}$ for steps. Thus, once the intercalation island
has nucleated, its discontinuity of expansion depends more strongly
on the denser network of Ir-steps. They play the role of a kinetic
barrier for diffusion of intercalated Cs atoms. An example of a tiling
step is highlighted in panels 5h and 5i of Fig. 5 where an arrow points
to an empty and filled tile, respectively.

\begin{figure}
\begin{centering}
\includegraphics{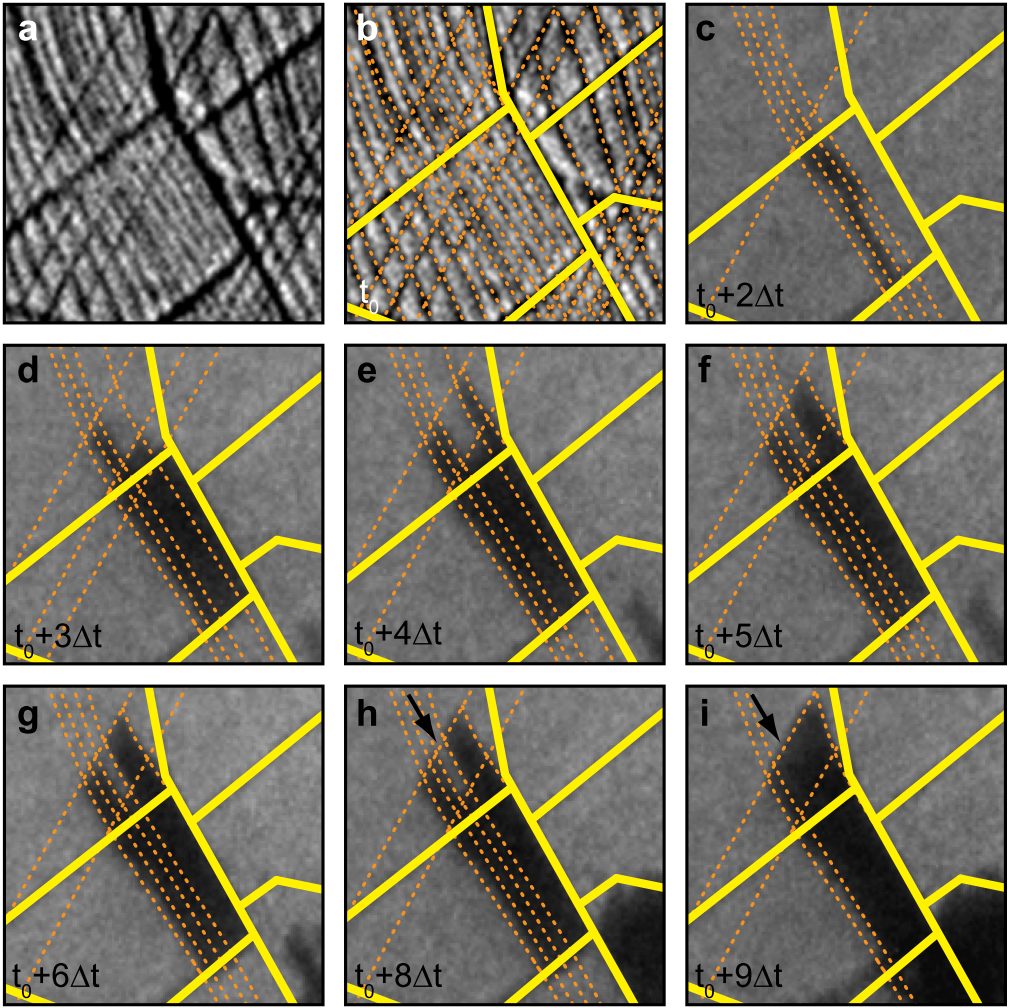}
\par\end{centering}

\caption{\textbf{Dynamics of intercalation.} Series of (2.52$\times$2.52)\textmu
m$^{2}$ LEEM frames showing the same surface area (\textbf{a},\textbf{b})
before any Cs deposition and (\textbf{c}-\textbf{i}) after the appearance
of the dark intercalated Cs-islands. Snapshots were imaged at the
indicated time intervals ($\Delta$\textit{t} = 72 s) with respect
to \textit{t}$_{0}$, the reference time when image \textbf{a} was
taken. In panel \textbf{b} the entire network of wrinkles and steps
is highlighted by full and dotted lines, respectively. In panels \textbf{c-i}
only wrinkles and relevant steps are highlighted. In the last two
panels an arrow indicates an empty (\textbf{h}) and a filled tile
(\textbf{i}).}

\end{figure}

As intercalated islands nucleate at crossing points of wrinkles, we
assume that the entrance points for Cs atoms to pass under graphene
are nano-scale cracks located at these crossing points, either as
pre-existing defects which pin the wrinkles—that is, their ends or
crossings—or as cracks which form during wrinkles formation due to
extensive local forces. As a wrinkle is a three-dimensional feature,
in LEEM it scatters electrons strongly and with poor coherence. This
contributes to lowering the resolution and prevents to resolve structural
inhomogeneities with LEEM.

\begin{figure}
\begin{centering}
\includegraphics{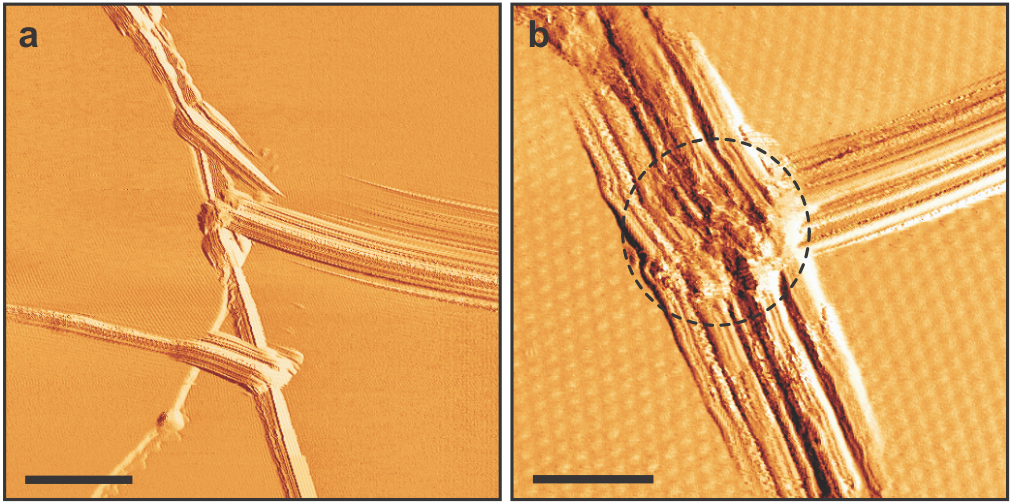}
\par\end{centering}

\caption{\textbf{STM} \textbf{characterization of wrinkles}. (\textbf{a}) A
large scale STM topograph showing several wrinkles. Scale bar corresponds
to 60 nm. (\textbf{b}) Area showing a crossing of two wrinkles. Scale
bar corresponds to 15 nm. The characteristic moiré structure of graphene
is visible in the areas around wrinkles. The crossing area highlighted
by the dotted circle appears as highly crumpled structure.}

\end{figure}

Our STM characterization of wrinkles of e-Gr/Ir in Fig. 6 enables
more insight than LEEM. The widths (10-50 nm) and heights (1-5 nm)
of different wrinkles vary. Most of the wrinkles orient independently
across terraces and step edges, whereas some of them are pinned near
to/at the step edges. Individual wrinkles can slightly change their
direction and sometimes also width or height (cf. Fig. 6a). In contrast
to the homogeneous parts, the wrinkle crossing shown in Fig. 6b depict
a highly inhomogeneous structure. Although it is not possible to resolve
cracks in such crumpled regions, their crumpling is consistent with
being the entry areas for Cs atoms. Less extensive defects have already
been characterized by STM in epitaxial graphene samples; Small graphene
rotations within the sheet resulted in an array of heptagon-pentagon
pairs \cite{Coraux2008} and larger rotations in nanometer-wide grain
boundaries \cite{Koepke2013}. Moreover, it was proposed that a vacancy
defect greater than four missing carbon atoms on a graphite surface
is needed to allow Cs intercalation to subsurface layers \cite{Buttner2011}.
Although our LEED data indicate uniform graphene, small rotations
(e.g. below 1\textdegree{}) and the existence of heptagon-pentagon
pairs cannot be entirely excluded. However, the LEEM and STM data
indicate that highly defective areas at wrinkles crossings enable
the transport of atoms across graphene layers. The role of these entry
cracks is additionally described in the following desorption experiment.

\begin{figure}
\begin{centering}
\includegraphics{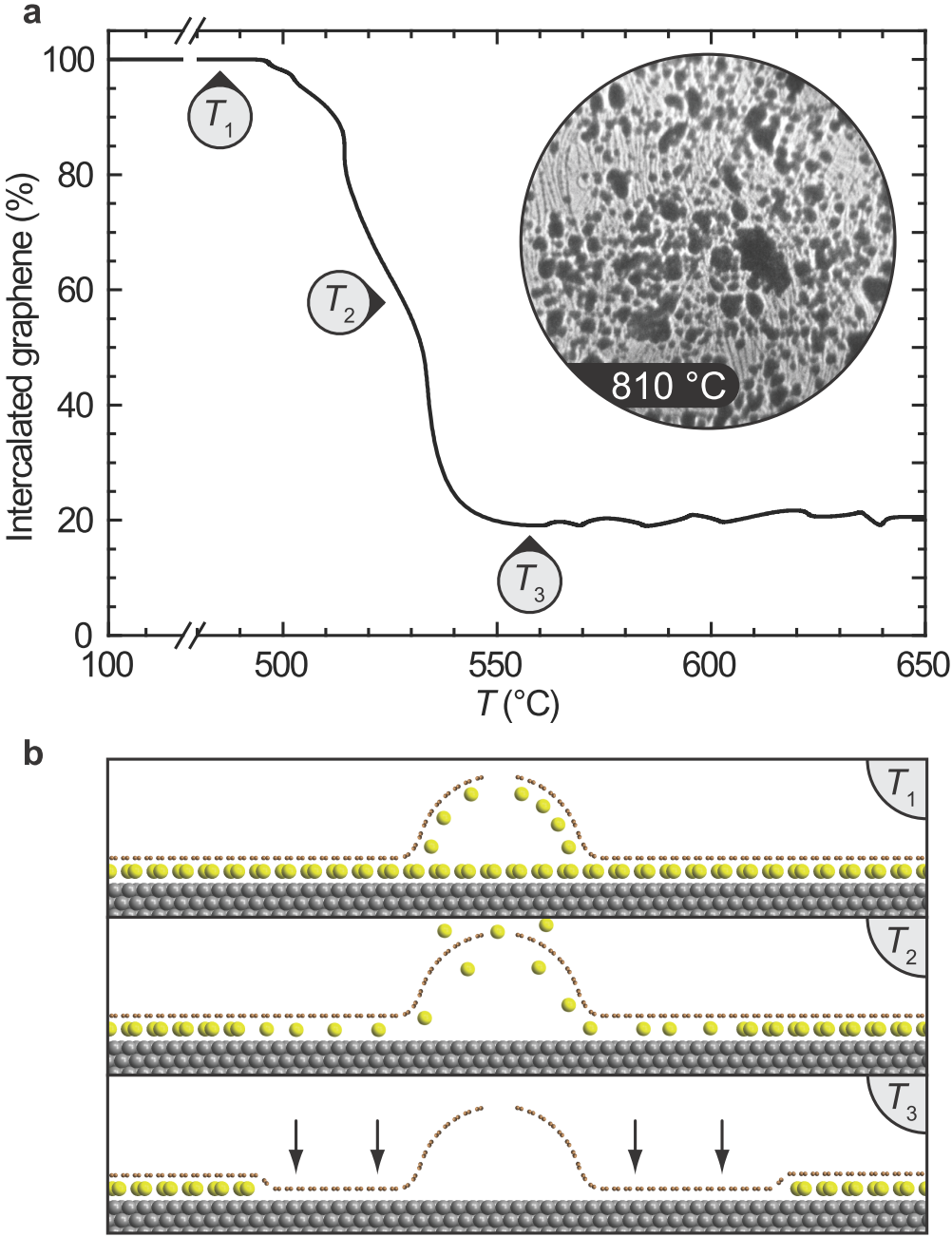}
\par\end{centering}

\caption{\textbf{Desorption of intercalated phase}. (\textbf{a}) Black curve
represents the \textgamma -phase LEEM contrast as a function of the
sample temperature, extracted by area integration from a LEEM movie.
The inset shows the final movie frame with well defined trapped intercalation
islands (image diameter is 14 \textmu m). (\textbf{b}) Schematic
sketch of the system structures at various temperatures: right before
desorption starts (\textit{T}$_{1}$), during Cs desorption (\textit{T}$_{2}$)
and after desorption is suppressed by relamination of graphene areas
(\textit{T}$_{3}$), as indicated by arrows.}

\end{figure}

Fig. 7a shows the analysis of the \textgamma -phase LEEM contrast
of the sample which was fully intercalated and then heated up to 810
\textdegree{}C. The entire desorption dynamics takes place in a narrow
temperature range (500-550 \textdegree{}C) where 80\% of the \textgamma
-phase contrast is lost and replaced by the contrast equivalent to
pristine e-Gr/Ir. However, the remaining 20\% of the \textgamma -phase
area does not vanish even at the highest recorded temperature; the
characteristic LEEM topograph (inset of Fig. 7a) shows well defi{}ned
dark pockets of the \textgamma -phase. We interpret the desorption
dynamics and its interruption as sketched in Fig. 7b. At about 500
\textdegree{}C desorption takes place, by the principle of micro-reversibility,
through the same channels that enabled intercalation—cracks in graphene
associated with wrinkles. When the Cs concentration falls below a
critical value, graphene will relaminate to Ir (cf. arrows in Fig.
7b lowermost panel). This will take place first in the surroundings
of the entry/exit spots where the Cs has already been lost. Once an
exit spot is surrounded entirely by relaminated e-Gr, no more desorption
takes place through this spot. Eventually, very stable Cs pockets
are trapped through the vdW adhesion of e-Gr to Ir(111).

\subsection{Li intercalation}

To make our study more comprehensive, we have characterized also the
intercalation of Li, which is on the opposite side of the alkali metals
group. Details are presented in Supplementary Fig. S6 and Supplementary
Note 5. Our van der Waals density functional (vdW-DF) calculations
show that intercalation is energetically more favourable even for
the diluted Li structure. This is confirmed by our ARPES data (cf.
Supplementary Fig. S6) which indicate the absence of phase separation
as observed for Cs. Homogeneous chemical doping uniformly increases
as Li is dosed and reaches its maximum when a ($1$\texttimes{}$1$)
structure relative to Ir is formed. This is possible because Li, due
to its small size, does not have to lift off graphene notably and
pay much for delamination energy in order to form intercalated structures.
Thereby it can intercalate at any density with much smaller cost of
graphene partial delamination. In the same manner, annealing of a
Li intercalated sample results in gradual and complete Li removal
indicating that no relamination effects take place.

\section{Discussion}

The observed kinetics of the system allows us to consider the mechanism
of the intercalation. First, in the \textalpha -phase the binding
of Cs-adatoms to graphene originates mainly from the partial delocalization
of the Cs s-electrons within graphene. However, the smaller the distance
between the adatoms gets, the larger is the Coulomb repulsion penalty.
Therefore the binding energy decreases with increasing Cs coverage
as shown by the blue data points in Fig. 4a; the repulsive adatom-adatom
interaction creates a two dimensional pressure that increases with
Cs coverage. The system relieves this pressure by emptying the adatom
phase through intercalation. This is the driving force for intercalation,
which takes place through entry-cracks at wrinkles crossings. Initial
intercalation is, however, prevented due to the energy cost of graphene
delamination from its substrate, about 50 meV per each C-atom \cite{Busse2011-1}.
The black-dashed line in Fig. 4a shows a fit to this value. As this
value is fixed per unit area, intercalation is only favorable if a
dense Cs-phase forms, where the delamination energy cost is shared
among many Cs atoms in the same small area. As the full lines in Fig.
4a indicate for a ($3$\texttimes{}$3$) and a denser structure, intercalation
is favored over an adatom phase of the same density. The additional
screening provided by the close proximity of the metallic substrate
in the intercalated phase lowers the Coulomb repulsion penalty for
delocalization and the corresponding energy gain overcompensates the
energy cost of graphene delamination. Experimentally, at the coverage
corresponding to a ($3$\texttimes{}$3$) structure, phase separation
of the system into an even denser intercalated \textgamma -phase
and a dilute adatom \textalpha -phase is observed, consistent with
the coverage dependence of the two phases as shown in Fig. 4a. In
the calculations where the vdW interaction is neglected, graphene
is essentially not bound to Ir(111). Thus in such an inadequate picture
no energetic penalty has to be paid for delamination, hence intercalated
phase is always preferred \cite{Lazic2012-1}. 

Although the vdW interaction is perceived as a weak dispersive force
with only minor effects in processes on surfaces dominated by chemical
bonds—for example, in catalysis—the vdW binding is quite generic.
This is best comprehended through the binding effect that scales with
the interacting surface: it is rather small for small molecules; however,
as a flat molecule increases in surface, the total amount of the vdW
force adds up. A good example is the relatively large phthalocyanine
molecule \cite{Brede2010} and the extreme limit is found in epitaxial
growth of graphene on surfaces. Moreover, if the bonding of a large
surface area of graphene competes with a more localized, and rather
strong chemical interaction—such as Cs intercalation—then both energy
scales may become comparable. In this case the semi-local DFT calculations
may even fail qualitatively in predicting surface structure. Without
the vdW interaction included, intercalation is always preferred due
to the better screening of Cs ions close to the Ir(111) substrate,
even for dilute structures, contrary to observation. This finding
highlights the importance of the vdW interactions when modelling and
understanding the intercalation/desorption process and points to the
importance of the delamination/relamination energy in the process
of transport of alkali atoms below graphene. 

Our findings are important for understanding the mechanism of intercalation
in layered systems where binding across the layers is determined by
the vdW interaction. The process of intercalation is the result of
a complex interplay of the adsorbed and intercalated phases: the intercalation
is triggered by the concentration of Cs that is large enough to overcome
the vdW binding of graphene to Ir. The precise energy threshold for
this process is determined by overcoming of the vdW binding of graphene
to Ir through the binding provided by the intercalated Cs atom. The
significance of the vdW energy parameter is much lower for Li, which
upon intercalation does not significantly change the separation between
graphene and Ir and thus can also form a dilute intercalation phase.
At the nano-scale, the process of nucleation of the intercalation
phase growth is guided by well-located structural features—that is,
nano-cracks at wrinkles crossings. The substrate steps are kinetic
barriers for the expansion of the intercalated layer. The intercalation
layer binds the floating graphene on the metal surface by electrostatic
forces. Finally, the vdW binding of graphene to the substrate is responsible
for the entrapment of isolated islands of intercalated Cs atoms, which
are chemically protected by graphene and thermally stable up to very
high temperatures.

\section{Methods}

\subsection{{\normalsize Graphene preparation}}

Ir(111) single crystals of the same type were used in all setups (99.99
\% purity and orientation accuracy better than 0.1\textdegree{}).
The substrate was cleaned by standard techniques of ion sputtering,
oxygen firing and annealing \cite{Kralj2011-1} and e-Gr was synthesized
by a combination of temperature programmed growth and chemical vapor
deposition procedures \cite{vanGastel2009-1}. This yields high-quality
e-Gr of uniform orientation as was confirmed after each preparation
by LEED. In ARPES setups, graphene quality was additionally verified
by spectral width (sharpness) of the Dirac cone bands. The sample
was always prepared \textit{in situ} in ultra-high vacuum (UHV) under
the base pressure better than 5\texttimes{}10$^{-8}$ (ARPES) and
5\texttimes{}10$^{-9}$ (LEEM and STM) Pa.

\subsection{{\normalsize Alkali metals deposition and desorption}}

Cs and Li were deposited by using commercially available alkali metal
dispensers (SAES Getters) with the typical flux values of $\sim$10$^{15}$
m$^{-2}$s$^{-1}$. During Cs deposition, the sample was kept at constant
temperature: 300 K in ARPES and STM setups and 340 K in LEEM setup.
During Li deposition the sample was kept at 300 K. In Cs desorption
experiments the LEEM data was recorded at an approximately constant
heating rate of 4 K s$^{-1}$ in the overall temperature range between
80 and 810 \textdegree{}C. Li desorption was performed by annealing
the sample in several steps up to the maximum temperature of 600 \textdegree{}C
with the ARPES data being recorded after each step.

\subsection{{\normalsize LEEM}}

Measurements were carried out in a commercial Elmitec LEEM III system
equipped with an energy analyzer, installed at beamline U5UA at NSLS
synchrotron. Data was recorded at 340 K (except for Fig. 4) with LEEM
system operating in bright-field mode. Imaging electron energy was
fixed to 1.5 eV (Cs deposition) or 2.4 eV (Cs desorption) while the
focus of the microscope was adjusted as required in order to optimize
the visibility of emerging structures. Spatial resolution of the microscope
was $\sim$10 nm.

\subsection{{\normalsize ARPES}}

Experiments were performed at U13UB beamline at NSLS synchrotron and
in ARPES dedicated setup in Zagreb. At U13UB, 23.4 eV p-polarized
light was used for excitation and spectra were recorded at 300 K with
Scienta SES2002 electron energy analyzer (10 meV energy and 0.14\textdegree{}
angular resolution) in the direction perpendicular to $\Gamma$K (Fig.
2, Supplementary Fig. S3 and S6). Spot size on the sample was (0.1\texttimes{}0.1)
mm$^{2}$. In Zagreb, a helium discharge lamp provided photons of
energy of 21.2 eV and of mixed polarization. Data was recorded with
Scienta SES100 electron energy analyzer (25 meV total energy resolution
and 0.2\textdegree{} angular resolution) at 300 K in the direction
parallel to $\Gamma$K (Supplementary Fig. S2). Spot diameter on the
sample in this setup was $\sim$2 mm.

\subsection{{\normalsize STM}}

STM measurements on intercalated system were performed at low-temperature
in Cologne (CreaTec LT-STM). The data was collected at 6 K, the STM
tip was virtually grounded and the sample was put to a bias voltage.
The data recorded in this setup is presented in Fig. 3a (1 V, 10 pA),
Fig. 3b (120 mV, 300 pA), Fig. 3c (40 mV, 300 pA) and Supplementary
Fig. S2b (1.07 V, 30 pA). The STM characterization of wrinkles was
performed in Zagreb (SPECS Aarhus STM). The data was collected at
room temperature, with the STM tip grounded and the sample put to
a bias voltage. In STM experiments wrinkles are sensitive to interactions
with the STM tip—that is, tip shape and tunneling parameters. The
data recorded in this setup is presented in Fig. 6a (780 mV, 750 pA)
and Fig. 6b (150 mV, 240 pA). STM data analysis was done in WSxM software
\cite{Horcas2007-1}.

\subsection{{\normalsize DFT calculations}}

DFT calculations were performed with the VASP code \cite{Kresse1993}
and the PAW implementation \cite{Klimes2011}. The plane wave energy
cutoff of 440 eV and a 15\texttimes{}15\texttimes{}1 \textit{k}-point
grid were used. The Ir substrate was simulated with a slab consisting
of five layers of iridium. A vacuum of 20 Å was provided in the \textit{z}-direction
separating the periodic slab images. The ionic positions were relaxed
until forces were below 0.01 eV Å$^{-1}$. The self-consistently implemented
vdW-DF \cite{Klimes2011,Dion2004-1} was used with opt-B88 as exchange
part \cite{Mittendorfer2011}. In the experiments, the e-Gr layer
on Ir(111) forms a $\sim$($10$\texttimes{}$10$)/($9$\texttimes{}$9$)
structure \cite{Busse2011-1}. In the calculations we have modified
the Ir lattice constant to make commensurate structures. The results
are in good agreement with large supercell calculations \cite{Brako2010}.
($2$\texttimes{}$2$), ($3$\texttimes{}$3$), ($4$\texttimes{}$4$)
and ($6$\texttimes{}$6$) in-plane unit cells of graphene were used
to calculate the corresponding Cs induced superstructures while ($1$\texttimes{}$1$)
and ($4$\texttimes{}$4$) in-plane unit cells were used in the case
of Li. All energy curves are calculated for the structures relaxed
with the vdW-DF functional—that is, both generalized gradient approximation
(GGA) \cite{Perdew1996} and vdW-DF calculations are done for the
same geometry of the system. For testing purposes we relaxed systems
containing Cs atoms with the vdW-DF and GGA functionals. The geometries
obtained are practically identical. This is not the case for the pristine
e-Gr/Ir system. By using GGA virtually no graphene binding results,
while vdW-DF yields a binding distance of 3.52 Å for a commensurate
cell. Charge density differences and non-local energy contributions
were obtained by making differences between the system and the system
parts, that is Ir slab, graphene sheet and a Cs atom—calculated at
coordinates they have in the relaxed system. Real space distribution
of the non-local binding energy density, as defined in Ref. \cite{Caciuc2012}
was obtained with the JuNoLo code \cite{Caciuc2012,Lazic2010,Callsen2012}.

\section*{Acknowledgements}

We acknowledge experimental assistance by F. Craes and J. Klinkhammer.
This work was supported by the Unity Through Knowledge Fund (grant
no. 66/10), the Deutsche Forschungsgemeinschaft (projects Bu2197/2
and INST 2156/514-1), the Ministry of Science and Technology of the
Republic of Croatia (contract no. 098-0352828-2863), and by the German
Academic Exchange Service \& Ministry of Science of the Republic of
Croatia via the project “Electrons in two dimensions”. Research carried
out at the Center for Functional Nanomaterials and National Synchrotron
Light Source, Brookhaven National Laboratory, is supported by the
U.S. Department of Energy, Office of Basic Energy Sciences, under
Contract No. DE-AC02-98CH10886.

\section*{Author contributions}

M.P. and M.K. conceived the research. M.K. supervised the research.
M.P., I.Š.R., J.T.S., M.M. and M.K. performed LEEM measurements. M.P.,
I.Š.R., I.P., Z.-H.P., M.M., P.P., T.V. and M.K. performed ARPES measurements.
M.P., S.R., C.B. and T.M. performed STM measurements. P.L., N.A.,
R.B and D.Š. performed DFT calculations. P.L. and N.A. provided the
discussion on the theoretical results. All authors discussed the data
and contributed to preparing the paper. M.P. prepared all figures.
T.V., T.M. and M.K. wrote the paper.

\section*{Additional information}

\textbf{Supplementary Information} accompanies this paper on http://www.nature.com/
naturecommunications 

\textbf{Competing fi{}nancial interests:} The authors declare no competing
fi{}nancial interests. 

\textbf{Reprints and permission} information is available online at
http://npg.nature.com/ reprintsandpermissions/ 

\textbf{How to cite this article:} Petrović, M. et al. The mechanism
of caesium intercalation of graphene. \textit{Nat. Commun.} 4:2772
DOI: 10.1038/ncomms3772 (2013).

\begin{thebibliography}{10}
\bibitem[1]{Emtsev2009-1}Emtsev, K. V. et al. Towards wafer-size
graphene layers by atmospheric pressure graphitization of silicon
carbide. \textit{Nature Mater.} \textbf{8,} 203–207 (2009).

\bibitem[2]{Wintterlin2009-1}Wintterlin, J. \& Bocquet, M.-L. Graphene
on metal surfaces. \textit{Surf. Sci.} \textbf{603,} 1841–1852 (2009).

\bibitem[3]{Li2009-1}Li, X. et al. Transfer of large-area graphene
fi{}lms for high-performance transparent conductive electrodes. \textit{Nano
Lett.} \textbf{9,} 4359–4363 (2009).

\bibitem[4]{Bae2010-1}Bae, S. et al. Roll-to-roll production of 30-inch
graphene fi{}lms for transparent electrodes. \textit{Nature Nanotech.}
\textbf{5,} 574–578 (2010).

\bibitem[5]{Lin2011-1}Lin, Y.-M. et al. Wafer-scale graphene integrated
circuit. \textit{Science} \textbf{332,} 1294–1297 (2011).

\bibitem[6]{Riedl2009-1}Riedl, C., Coletti, C., Iwasaki, T., Zakharov,
A. A. \& Starke, U. Quasi-free-standing epitaxial graphene on SiC
obtained by hydrogen intercalation. \textit{Phys. Rev. Lett.} \textbf{103,}
246804 (2009).

\bibitem[7]{Varykhalov2008-1}Varykhalov, A. et al. Electronic and
magnetic properties of quasifreestanding graphene on Ni. \textit{Phys.
Rev. Lett.} \textbf{101,} 157601 (2008).

\bibitem[8]{Liu2011}Liu, H., Liu, Y. \& Zhu, D. Chemical doping of
graphene. \textit{J. Mater. Chem.} \textbf{21}, 3335 (2011).

\bibitem[9]{Novoselov2004-1}Novoselov, K. S. et al. Electric fi{}eld
effect in atomically thin carbon fi{}lms. \textit{Science} \textbf{306,}
666–669 (2004).

\bibitem[10]{Jung2009-1}Jung, N. et al. Charge transfer chemical
doping of few layer graphenes: charge distribution and band gap formation.
\textit{Nano Lett.} \textbf{9,} 4133–4137 (2009).

\bibitem[11]{McChesney2010-1}McChesney, J. L. et al. Extended van
Hove singularity and superconducting instability in doped graphene.
\textit{Phys. Rev. Lett.} \textbf{104,} 136803 (2010).

\bibitem[12]{Dresselhaus1981}Dresselhaus, M. S. \& Dresselhaus, G.
Intercalation compounds of graphite. \textit{Adv. Phys.} \textbf{30},
139–326 (1981).

\bibitem[13]{Chung2002}Chung, D. D. L. Review Graphite. \textit{J.
Mater. Sci.} \textbf{37}, 1–15 (2002).

\bibitem[14]{Valla2012}Valla, T., Pan, Z.-H., Gardner, D., Lee, Y.
S. \& Chu, S. Photoemission spectroscopy of magnetic and nonmagnetic
impurities on the surface of the Bi$_{2}$Se$_{3}$ topological insulator.
\textit{Phys. Rev. Lett.} \textbf{108}, 117601 (2012).

\bibitem[15]{Weller2005-1}Weller, T. E., Ellerby, M., Saxena, S.
S., Smith, R. P. \& Skipper, N. T. Superconductivity in the intercalated
graphite compounds C$_{6}$Yb and C$_{6}$Ca. \textit{Nature Phys.}
\textbf{1,} 39–41 (2005).

\bibitem[16]{Emery2005-1}Emery, N. et al. Superconductivity of bulk
CaC$_{6}$. \textit{Phys. Rev. Lett.} \textbf{95,} 087003 (2005).

\bibitem[17]{Profeta2012-1}Profeta, G., Calandra, M. \& Mauri, F.
Phonon-mediated superconductivity in graphene by lithium deposition.
\textit{Nature Phys.} \textbf{8,} 131–134 (2012).

\bibitem[18]{Uchoa2007-1}Uchoa, B. \& Castro Neto, A. H. Superconducting
states of pure and doped graphene. \textit{Phys. Rev. Lett.} \textbf{98,}
146801 (2007).

\bibitem[19]{Yoo2011-1}Yoo, J. J. et al. Ultrathin planar graphene
supercapacitors. \textit{Nano Lett.} \textbf{11,} 1423–1427 (2011).

\bibitem[20]{Emtsev2011}Emtsev, K., Zakharov, A., Coletti, C., Forti,
S. \& Starke, U. Ambipolar doping in quasifree epitaxial graphene
on SiC(0001) controlled by Ge intercalation. \textit{Phys. Rev. B}
\textbf{84}, 125423 (2011).

\bibitem[21]{Sicot2012-1}Sicot, M. et al. Size-selected epitaxial
nanoislands underneath graphene moiré on Rh(111). \textit{ACS Nano}
\textbf{6,} 151–158 (2012).

\bibitem[22]{Decker2013}Decker, R. et al. Atomic-scale magnetism
of cobalt-intercalated graphene. \textit{Phys. Rev. B} \textbf{87},
041403 (2013).

\bibitem[23]{Pacile2013}Pacilé, D. et al. Artificially lattice-mismatched
graphene/metal interface: Graphene/Ni/Ir(111). \textit{Phys. Rev.
B} \textbf{87}, 035420 (2013).

\bibitem[24]{Schumacher2013-1}Schumacher, S., Förster, D. F., Rösner,
M., Wehling, T. O. \& Michely, T. Strain in epitaxial graphene visualized
by intercalation. \textit{Phys. Rev. Lett.} \textbf{110,} 086111 (2013).

\bibitem[25]{vanGastel2009-1}van Gastel, R. et al. Selecting a single
orientation for millimeter sized graphene sheets. \textit{Appl. Phys.
Lett.} \textbf{95,} 121901 (2009).

\bibitem[26]{Busse2011-1}Busse, C. et al. Graphene on Ir(111): Physisorption
with chemical modulation. \textit{Phys. Rev. Lett.} \textbf{107,}
036101 (2011).

\bibitem[27]{Pletikosic2009-1}Pletikosić, I. et al. Dirac cones and
minigaps for graphene on Ir(111). \textit{Phys. Rev. Lett.} \textbf{102,}
056808 (2009).

\bibitem[28]{Kralj2011-1}Kralj, M. et al. Graphene on Ir(111) characterized
by angle-resolved photoemission. \textit{Phys. Rev. B} \textbf{84,}
075427 (2011).

\bibitem[29]{Bianchi2010-1}Bianchi, M. et al. Electron-phonon coupling
in potassium-doped graphene: Angle-resolved photoemission spectroscopy.
\textit{Phys. Rev. B} \textbf{81,} 041403 (2010).

\bibitem[30]{Pletikosic2012-1}Pletikosić, I., Kralj, M., Milun, M.
\& Pervan, P. Finding the bare band: Electron coupling to two phonon
modes in potassium-doped graphene on Ir(111). \textit{Phys. Rev. B}
\textbf{85,} 155447 (2012).

\bibitem[31]{Tontegode1991-1}Tontegode, A. Carbon on transition metal
surfaces. \textit{Prog. Surf. Sci.} \textbf{38,} 201–429 (1991).

\bibitem[32]{Granas2012-1}Grånäs, E. et al. Oxygen intercalation
under graphene on Ir(111): Energetics, kinetics, and the role of graphene
edges. \textit{ACS Nano} \textbf{6,} 9951–9963 (2012).

\bibitem[33]{Larciprete2012-1}Larciprete, R. et al. Oxygen switching
of the epitaxial graphene–metal interaction. \textit{ACS Nano} \textbf{6,}
9551–9558 (2012).

\bibitem[34]{Rutkov1995-1}Rut’kov, E., Tontegode, A. \& Usufov, M.
Evidence for a C$_{60}$ monolayer intercalated between a graphite
monolayer and iridium. \textit{Phys. Rev. Lett.} \textbf{74,} 758–760
(1995).

\bibitem[35]{Vinogradov2013-1}Vinogradov, N. A. et al. Hole doping
of graphene supported on Ir(111) by AlBr$_{3}$. \textit{Appl. Phys.
Lett.} \textbf{102,} 061601 (2013).

\bibitem[36]{Hattab2012-1}Hattab, H. et al. Interplay of wrinkles,
strain, and lattice parameter in graphene on iridium. \textit{Nano
Lett.} \textbf{12,} 678–682 (2012).

\bibitem[37]{NDiaye2009-1}N’Diaye, A. T. et al. In situ observation
of stress relaxation in epitaxial graphene. \textit{New J. Phys.}
\textbf{11,} 113056 (2009).

\bibitem[38]{Chung2003}Chung, W. et al. Layer Spacings in Coherently
Strained Epitaxial Metal Films. \textit{Phys. Rev. Lett.} \textbf{90},
216105 (2003).

\bibitem[39]{Coraux2008}Coraux, J., N’Diaye, A. T., Busse, C. \&
Michely, T. Structural coherency of graphene on Ir(111). \textit{Nano
Lett.} \textbf{8}, 565–70 (2008).

\bibitem[40]{Koepke2013}Koepke, J. C. et al. Atomic-scale evidence
for potential barriers and strong carrier scattering at graphene grain
boundaries: a scanning tunneling microscopy study. \textit{ACS Nano}
\textbf{7}, 75–86 (2013).

\bibitem[41]{Buttner2011}Büttner, M., Choudhury, P., Karl Johnson,
J. \& Yates, J. T. Vacancy clusters as entry ports for cesium intercalation
in graphite. \textit{Carbon} \textbf{49}, 3937–3952 (2011).

\bibitem[42]{Lazic2012-1}Lazić, P. et al. Rationale for switching
to nonlocal functionals in density functional theory. \textit{J. Phys.:
Condens. Matter} \textbf{24,} 424215 (2012).

\bibitem[43]{Brede2010}Brede, J. et al. Spin- and Energy-Dependent
Tunneling through a Single Molecule with Intramolecular Spatial Resolution.
\textit{Phys. Rev. Lett.} \textbf{105}, 047204 (2010).

\bibitem[44]{Horcas2007-1}Horcas, I. et al. WSXM: a software for
scanning probe microscopy and a tool for nanotechnology. \textit{Rev.
Sci. Instrum.} \textbf{78,} 013705 (2007).

\bibitem[45]{Kresse1993}Kresse, K. \& Hafner, J. Ab initio molecular
dynamics for liquid metals. \textit{Phys. Rev. B} \textbf{47}, 558
(1993).

\bibitem[46]{Klimes2011}Klimeš, J., Bowler, D. \& Michaelides, A.
Van der Waals density functionals applied to solids. \textit{Phys.
Rev. B} \textbf{83}, 195131 (2011).

\bibitem[47]{Dion2004-1}Dion, M., Rydberg, H., Schröder, E., Langreth,
D. C. \& Lundqvist, B. I. Van der Waals density functional for general
geometries. \textit{Phys. Rev. Lett.} \textbf{92,} 246401 (2004).

\bibitem[48]{Mittendorfer2011}Mittendorfer, F. et al. Graphene on
Ni(111): Strong interaction and weak adsorption. \textit{Phys. Rev.
B} \textbf{84}, 201401 (2011).

\bibitem[49]{Brako2010}Brako, R., Šokčević, D., Lazić, P. \& Atodiresei,
N. Graphene on the Ir(111) surface: from van der Waals to strong bonding.
\textit{New J. Phys.} \textbf{12}, 113016 (2010).

\bibitem[50]{Perdew1996}Perdew, J. P., Burke, K. \& Ernzerhof, M.
Generalized Gradient Approximation Made Simple. \textit{Phys. Rev.
Lett.} \textbf{77}, 3865–3868 (1996).

\bibitem[51]{Caciuc2012}Caciuc, V., Atodiresei, N., Callsen, M.,
Lazić, P. \& Blügel, S. Ab initio and semi-empirical van der Waals
study of graphene–boron nitride interaction from a molecular point
of view. \textit{J. Phys.: Condens. Matter} \textbf{24}, 424214 (2012).

\bibitem[52]{Lazic2010}Lazić, P. et al. JuNoLo—Jülich nonlocal code
for parallel post-processing evaluation of vdW-DF correlation energy.
\textit{Comput. Phys. Commun.} \textbf{181}, 371–379 (2010).

\bibitem[53]{Callsen2012}Callsen, M., Atodiresei, N., Caciuc, V.
\& Blügel, S. Semiempirical van der Waals interactions versus ab initio
nonlocal correlation effects in the thiophene-Cu(111) system. \textit{Phys.
Rev. B} \textbf{86}, 085439 (2012).

\end{thebibliography}
\end{document}